\documentclass[draft]{aipproc}
\layoutstyle{6x9}
\usepackage{epsfig}
\usepackage{amssymb}

\begin{document}
\title{ The Strange Quark Polarisation from Charged Kaon
Production on Deuterons  
}


\author{R.Windmolders
 (On behalf of the COMPASS collaboration)}{
address ={Universit\"{a}t Bonn, Physikalisches Institut, 53115 Bonn, Germany}  
}

\begin{abstract}
The  strange quark helicity distribution  $\Delta s(x)$ is
derived at LO from the semi-inclusive and inclusive spin asymmetries 
 measured by the COMPASS experiment at
CERN.
The significance of the results is found to depend critically  on
the ratio of the ${\overline s}$ and $u$ quark fragmentation functions
into kaons
$\int D_{\overline s}^{K+}(z)dz/\int D_u^{K+}(z)dz$.

\end{abstract}

\maketitle
In contrast to other sea  quarks, 
 the strange quark  contribution to the nucleon spin is  accessible in
inclusive lepton-nucleon scattering experiments.
 Already twenty years ago
the EMC \cite{emc} has found
 its first moment 
$\Delta s + \Delta {\overline s}$  to be negative 
under the assumption of SU(3)$_F$ symmetry in hyperon $\beta$ decays
 and this result has been confirmed
with improved precision by recent measurements performed by
HERMES \cite{hrm2007}
($\Delta s + \Delta {\overline s} = -0.103 \pm 0.007 {\rm(exp.)} \pm 0.013 {\rm
(theor.)} \pm 0.008 {\rm (evol.)}$) 
and by COMPASS 
 \cite{dval}: 
\begin{equation}
\Delta s + \Delta {\overline s} = -0.09 \pm 0.01 {\rm(stat.)} \pm 0.02 {\rm
(syst.)}.
\end{equation}
Inclusive experiments   provide a direct evaluation of the first moment
of $(\Delta s+ \Delta {\overline s})$ only. However
 the distribution $ \Delta s(x,Q^2)$ can be
obtained from semi-inclusive channels in which  interactions on
strange quarks are favoured, such as charged kaon production. These
measurements  require
final state particle identification and  became only feasible in recent
experiments  \cite{hrm2005,hrm2008}.

In this paper we present a leading order (LO) evaluation of the polarised parton distributions
 $\Delta u_v + \Delta d_v$, $\Delta {\overline u} + \Delta {\overline d}$
 and $\Delta s (= \Delta {\overline s})$ derived from the inclusive and
semi-inclusive spin asymmetries measured by the COMPASS experiment.
The data sample covers the years 2002-2004 and 2006, and corresponds to the full
statistics collected on the longitudinally polarised $^6$LiD target exposed
to the 160 GeV muon beam at CERN.
The averaged beam and target polarisations were $- 0.80$ and $0.50$, respectively.
 Each selected event contains a beam muon with energy $140 < E_{\mu} < 180$ GeV,
  a scattered muon and, for the semi-inclusive samples, at least
one charged hadron originating from the interaction vertex. 
The DIS conditions
$Q^2 \ge 1 ({\rm GeV}/c)^2$ and $0.1 \le y \le 0.9$ are required in the
muon kinematics. The hadron is required to carry a fractional energy
$0.20 < z < 0.85$ and must be identified as a pion or a kaon by the RICH detector,
which limits its momentum to the range $10 < p < 50$ GeV/c.
The total number of identified hadrons is $23 \cdot 10^6$ $\pi^+$, $21 \cdot 10^6$ $\pi^-$,
$4.8 \cdot 10^6$ $K^+$ and $3.3 \cdot 10^6$ $K^-$.
 The purity of
the selected samples is above 95 \% for the pions and between 80 and
90 \% for the kaons.  

\begin{figure}[here]
\centering
\epsfig{file=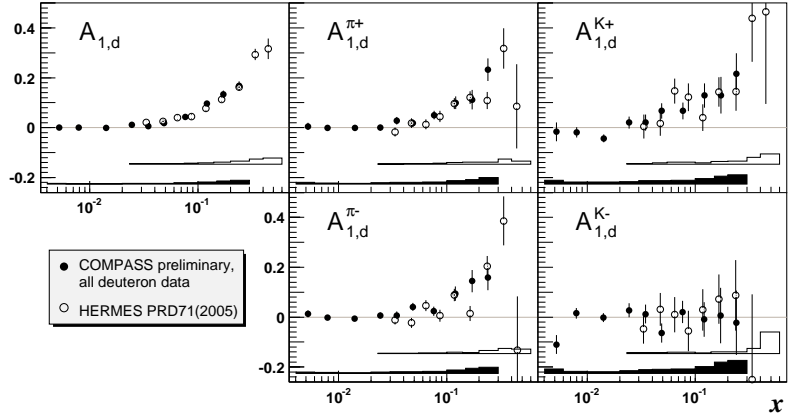,width=0.7\textwidth}
\caption{Spin asymmetries of COMPASS compared to the HERMES results \cite{hrm2005}.
  The bands show the systematic uncertainties for the two sets of data.
}
\label{fig:Asym_with_HERMES}
\end{figure}

The inclusive and semi-inclusive asymmetries are shown 
in Fig.\,\ref{fig:Asym_with_HERMES}
 as a function of the scaling variable $x$ in a range
limited by the kinematical threshold $Q^2 \ge 1 ({\rm GeV}/c)^2$,
which implies $x > 0.004$, 
 and a value where
sea quarks distributions become insignificant
 $( x < 0.3)$.\\

 The  correction
applied to account for the contamination of the pion and kaon samples
has only a small effect on the 
semi-inclusive asymmetries. 
The HERMES data \cite{hrm2005}, shown on the same plots,
 are in good agreement with the present ones in the region of overlap and
their quoted errors are generally also comparable. The COMPASS systematic
error includes an overall scale uncertainty  due to the uncertainties
on the beam and target polarisation, on the dilution factor and on the
depolarisation factor which  amount to 8 \% when added in quadrature.
In addition, each point has  an uncertainty due to possible apparatus
induced asymmetries which are estimated to be less than 0.4 $\sigma({\rm stat}.)$.


The  polarised parton densities (PDFs) have been evaluated at $Q^2_0=3 ({\rm GeV}/c)^2$  
in each bin of $x$ 
by a least square fit on the  asymmetries assumed to be independent of $Q^2$ (Fig.\,2).
In this calculation, unpolarised PDFs are taken from the MRST parameterisation
 \cite{mrst} and fragmentation functions (FFs) from the recent DSS parameterisation \cite{flo_dss},
both at LO. 
In order to test the dependence of the polarised PDFs on the FFs, we also show
the values obtained with the EMC FFs  \cite{emc_ff}. In 
contrast to other parameterisations which are derived from global fits, the
latter ones have been  extracted from the EMC data only, so that 
only the $u$ quark fragmentation
could be measured. Therefore, in addition to the usual assumptions made to reduce the
number of FFs, it was also assumed that $D_{\overline s}^{K+} =  D_u^{\pi+}$.

The comparison of the two sets of values shows that the valence and non-strange
sea distributions depend very little on the choice of FFs while  $\Delta s(x)$
and its statistical error increase by a factor of  2-3  when the
DSS FFs are replaced by the  EMC ones. The
valence  distribution is in good agreement with the DNS
parameterisation \cite{DNS} and with a previous evaluation based on the difference asymmetry
for non-identified hadrons $A^{h^+-h^-}$ \cite{dval}:
 its first moment truncated to
 the measured range  is $0.28 \pm 0.06 ({\rm stat.}) \pm 0.03 ({\rm syst.})$, vs.
 $0.26 \pm 0.07 \pm 0.04 $ at $Q^2 = 10$ (GeV/c)$^2$ in
Ref.\,\cite{dval}.
The non-strange sea distribution is consistent with zero for all values of $x$.
The strange quark distribution obtained with the DSS FFs is also consistent with zero
and its truncated moment is $-0.01 \pm 0.01 \pm 0.01$ (vs. $-0.05 \pm 0.03 \pm 0.01$ with the
EMC FFs).\\

\begin{figure}[here]
\centering
\includegraphics[width=0.5\textwidth,clip]{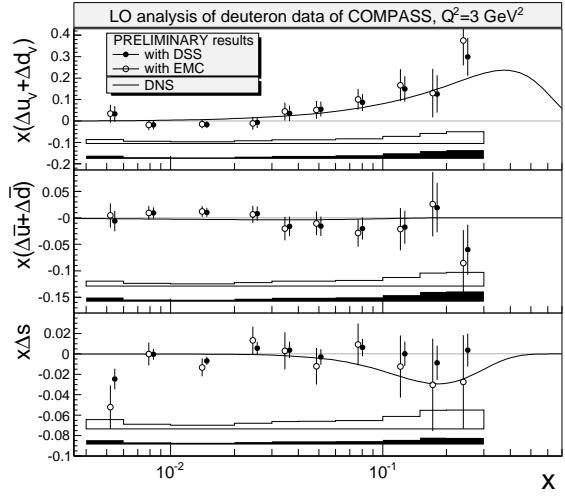}
\caption{
The quark helicity distributions evaluated at  $Q^2_0=3 ({\rm GeV}/c)^2$
  as a function of $x$ for two sets of fragmentation functions (DSS
\cite{flo_dss} and EMC\cite{emc_ff}).
  The bands show the corresponding systematic uncertainties.
  The curves represent the LO DNS parameterisation of polarised PDFs \cite{DNS}.
}
\label{fig:PDF_Syst_EMC_DNS}
\end{figure}
The dependence of $\Delta s(x)$ on the FFs can be further explored by introducing the
charged kaon asymmetry 
\begin{equation}
A^{K^++K^-} = \Bigl [\sigma^{K^+} A^{K^+} + \sigma^{K^-} A^{K^-} \Bigr ] /
\Bigl [\sigma^{K^+} + \sigma^{K^-} \Bigr ].
\end{equation}
 These asymmetries  are found to be very stable  
with respect to  changes in the cross-section ratio $\sigma^{K^-}/\sigma^{K^+}$.
At LO this ratio  only depends on  
unpolarised PDFs and on the ratios of  
unfavoured to favoured and strange to favoured FFs:
$R_{UF} = \int D_d^{K^+}(z) dz/ \int D_u^{K^+}(z) dz $ and
$R_{SF} = \int D_{\overline s}^{K^+}(z) dz/ \int D_u^{K^+}(z) dz $,
which are equal to 0.14 and 6.6 for the DSS FFs at $Q^2 = 3 ({\rm GeV}/c)^2$
(0.35 and 3.4 for the EMC FFs).
The values shown in Fig.\,(3, left) have been  obtained
with the MRST PDFs and the DSS FFs. 
  As for the $K^+$ and $K^-$
asymmetries, they are in  very good agreement with the HERMES values of Ref.\,\cite{hrm2005}. 
  
The strange quark polarisation is related to $ A^{K^++K^-}$ and to 
the inclusive asymmetry $A_1^d$ by the linear relation 
\begin{equation}
\frac {\Delta s}{s} = A_1^d + (A^{K^++K^-}\!\! - A_1^d) \frac{Q/s + \alpha}{\alpha - 0.8}
\end{equation}
where $ Q = u + {\overline u} + d + {\overline d} $ is the non-strange quark density and
 $\alpha = (2 R_{UF} + 2 R_{SF})/(3 R_{UF} + 2) $. 
As expected, the use of the above formula with the $A^{K^++K^-}$ values of Fig.\,3 leads to values 
of $\Delta s$ practically equal to those of Fig.\,2, with slightly larger statistical errors.
Several other interesting features can be derived from Eq.(3).
In the  special case where $A^{K^++K^-} = A_1^d$, $\Delta s$ becomes insensitive to the
FFs and its first moment is  very small and positive. Otherwise the main dependence on the FFs is
due to $R_{SF}$ which appears only in the numerator of $\alpha$, and its effect is  
amplified by the large values of the ratio $Q/s$.  
At low $x$ where $A_1^d \approx 0$, negative
values of $A^{K^++K^-}$ imply $\Delta s < 0$. The COMPASS values of $A^{K^++K^-}$ 
at low $x$ provide at least
a hint that $\Delta s$ may become negative 
in the previously unmeasured region  $x <0.02$, as predicted in the recent
DSSV fit \cite{florian2008}. \\

\begin{figure}[here]
\centering
\includegraphics[width=0.45\textwidth,clip]{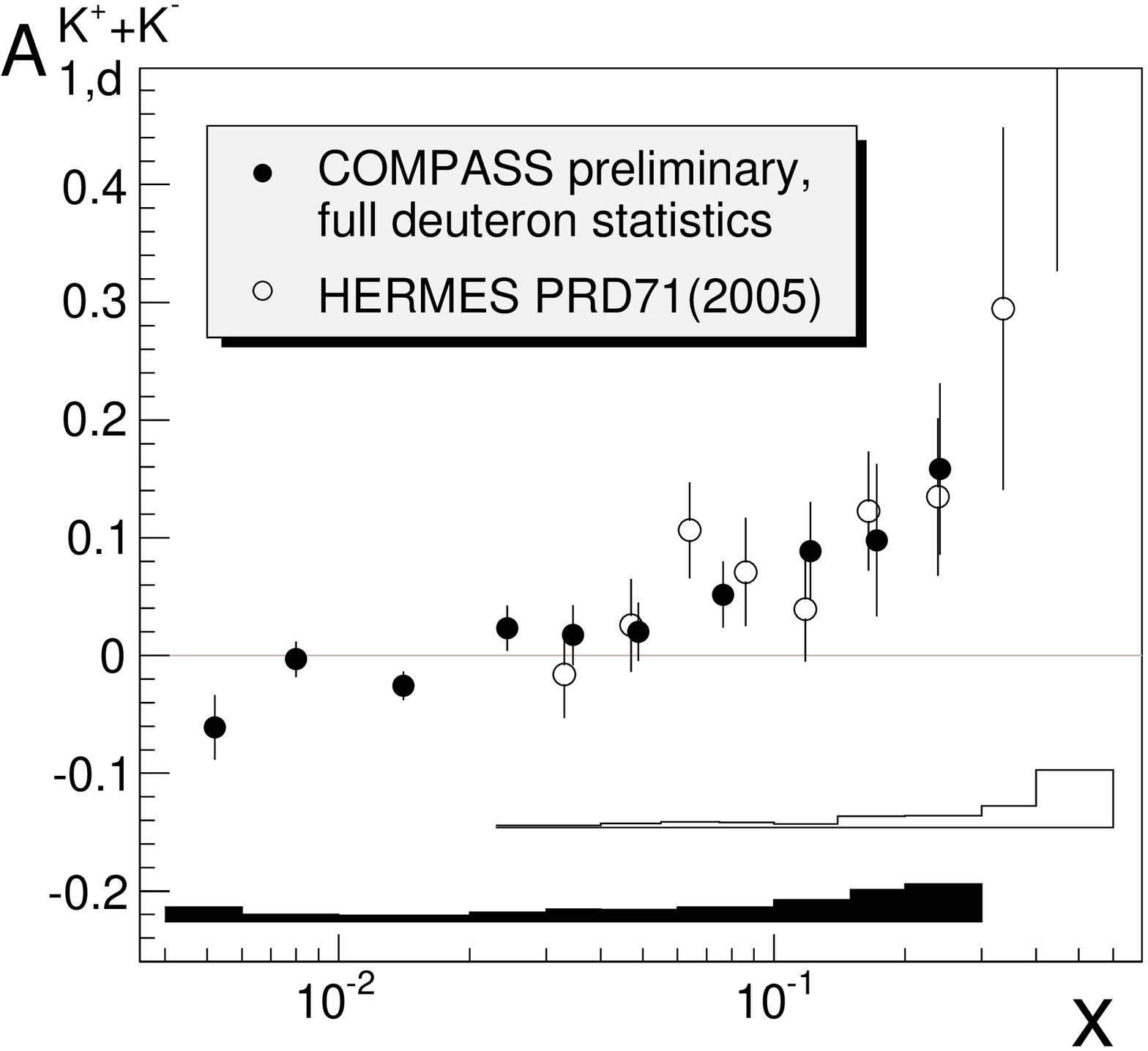}
\hfill
\includegraphics[width=0.60\textwidth,clip]{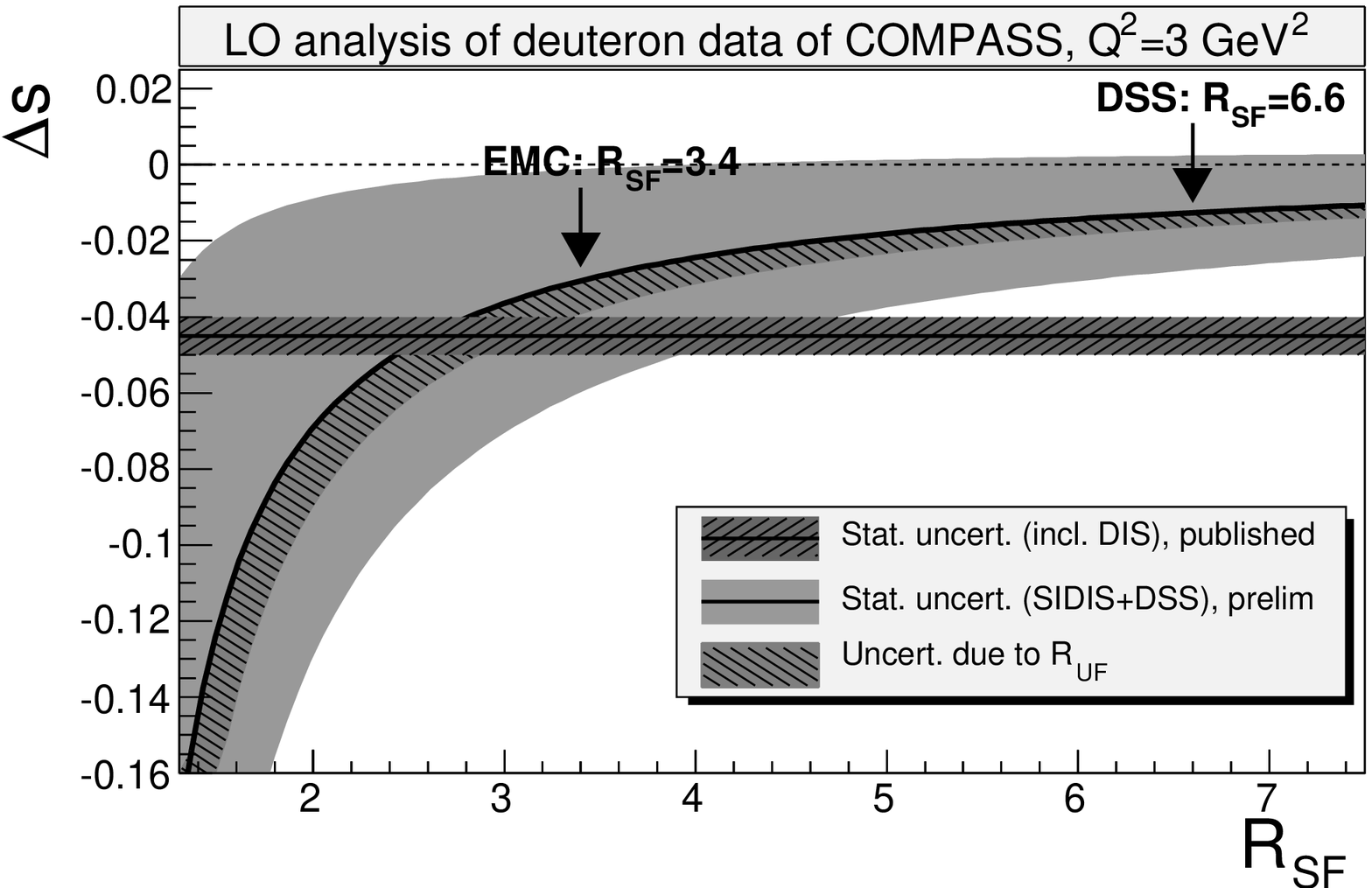}
\caption{
{\it Left:}
Charged kaon asymmetries obtained with cross-section weights
  from MRST PDF's and DSS FF's. 
{\it Right:}
Integral of $\Delta s$ over the measured range of $x$, as a function of the ratio
  $R_{SF}$ for $R_{UF}$ fixed at the DSS value of 0.14 (thick solid curve).
  The light-gray area shows the statistical uncertainty and the hatched band
  inside of it shows the effect of increasing $R_{UF}$ to 0.35 (EMC value).
  The horizontal band represents the full moment of $\Delta s$ derived from
  the COMPASS value of $\Gamma_1^N$ (Eq.\,1).
}
\label{fig:kpm-xds-integ2}
\end{figure}
Fig.\,(3, right) shows the variation of the first moment of $\Delta s$ 
truncated to the measured region as
a function of $R_{SF}$. For  $R_{SF} \gtrsim 5$, we observe that the values are close to zero and
larger than the full moment derived from the inclusive analysis (Eq.1). In particular this is
the case for the DSS FFs where  $R_{SF} = 6.6$. The difference never exceeds two 
standard deviations so that
no firm conclusions can be drawn from the COMPASS data alone but, as shown on Fig.\,(3, left) the
HERMES data lead to a similar result. In contrast, if  $R_{SF} \lesssim 4$, $A^{K^++K^-}$
becomes less and less sensitive to $\Delta s$ because $D_{\overline s}^{K^+}$ is small.  \\
The conclusions on $\Delta s$ obtained in the present analysis are thus conditional
 on the ratio $R_{SF}$ which will later 
 be derived from the $K^+$ and $K^-$ rates 
observed in the COMPASS data.


\end{document}